\documentclass[conference]{IEEEtran}
\IEEEoverridecommandlockouts
\usepackage{cite}
\usepackage{amsmath,amssymb,amsfonts}
\usepackage{algorithmic}
\usepackage{graphicx}
\usepackage{textcomp}
\usepackage{xcolor}
\usepackage{listings}
\PassOptionsToPackage{hyphens}{url}
\usepackage[hidelinks]{hyperref}
\def\BibTeX{{\rm B\kern-.05em{\sc i\kern-.025em b}\kern-.08em
    T\kern-.1667em\lower.7ex\hbox{E}\kern-.125emX}}
\begin{document}

\title{Exploration on Real World Assets (RWAs) \& Tokenization\\}

\author{\IEEEauthorblockN{Ning Xia}
\IEEEauthorblockA{nx2173@columbia.edu}
\and
\IEEEauthorblockN{Xiaolei Zhao}
\IEEEauthorblockA{xz3283@columbia.edu}
\and
\IEEEauthorblockN{Yimin Yang}
\IEEEauthorblockA{yy3352@columbia.edu}
\and
\IEEEauthorblockN{Yixuan Li}
\IEEEauthorblockA{yl5468@columbia.edu}
\and
\IEEEauthorblockN{Yucong Li}
\IEEEauthorblockA{yl5363@columbia.edu}
}
\pagestyle{plain}
\maketitle

\begin{abstract}
This study delves into the tokenization of real-world assets (RWAs) on the blockchain with the objective of augmenting liquidity and refining asset management practices. By conducting an exhaustive analysis of the technical procedures implicated and scrutinizing case studies of existing deployments, this research evaluates the advantages, hurdles, and prospective advancements of blockchain technology in reshaping conventional asset management paradigms.
\end{abstract}

\section{Introduction}
Real-world assets (RWAs) serve as the bedrock of the traditional financial system, encompassing tangible assets like real estate, commodities, and securities. These assets underpin investment portfolios, offering stability and diversification within the market landscape.

Blockchain technology, with its decentralized ledger system, has emerged as a disruptive force in asset management. Its hallmark attributes of transparency, security, and efficiency have paved the way for asset tokenization—where physical assets are digitized into tokens on blockchain networks. This innovation holds the promise of revolutionizing how assets are managed and traded, unlocking new levels of liquidity and accessibility within financial markets.

The real estate industry stands on the cusp of a digital revolution, driven by the convergence of blockchain technology and property ownership. Real estate tokenization, the process of representing property ownership as digital tokens on a blockchain, offers a paradigm shift in how real estate assets are managed, traded, and accessed. In this paper, we delve into the intricate details of real estate tokenization, dissecting its technical underpinnings, financial implications, and operational intricacies. By examining the blockchain infrastructure, token standards, compliance frameworks, and market dynamics associated with real estate tokenization, we aim to provide a comprehensive understanding of its potential impact on the real estate market. Through case studies, industry insights, and future projections, we explore the transformative power of real estate tokenization in democratizing access to real estate investment, enhancing market liquidity, and redefining asset management practices.

\section{Background and Motivation}
 Before delving into the potential of blockchain solutions for RWAs, it's essential to understand the constraints inherent in traditional asset management practices and the existing blockchain solutions for RWAs.
 
\subsection{Limitations of Traditional Asset Management}
In traditional asset management, the process of trading and investing entails the exchange of capital within a framework of strict regulations. This process faces various limitations, such as high entry barriers, restricted liquidity, and indivisible ownership.

From the perspective of investors and consumers, many investment opportunities require substantial amounts of capital to participate. This high upfront cost can make it difficult for individuals with limited financial resources to access these investment opportunities. For example, the initial cost of purchasing property is very high, so individuals who do not have an accumulation of capital are more likely to miss the chance of investing in real estate.

In addition, these types of investments are regulated and may necessitate individuals to adhere to certain regulatory criteria before participating. These rules and regulations not only prolong the investment process but also exclude many potential qualified investors and consumers.

From the perspective of owners, the only way to turn their physical assets into cash is to sell or lease the entire assets to other individuals. Yet, this procedure can be daunting because it may take months or even years for the owner to find suitable buyers or tenants at the desired price. The lengthy process involves negotiations, auctions, and so on. It's not feasible to split ownership of the asset, selling a portion while retaining some for themselves \cite{b1}. For instance, the owner cannot sell only a portion of his/her real estate or artwork. This situation underscores the illiquidity of RWAs.

\subsection{Existing Blockchain Solutions for RWAs}
\begin{figure}[htbp]
\center
\includegraphics[width=\columnwidth]{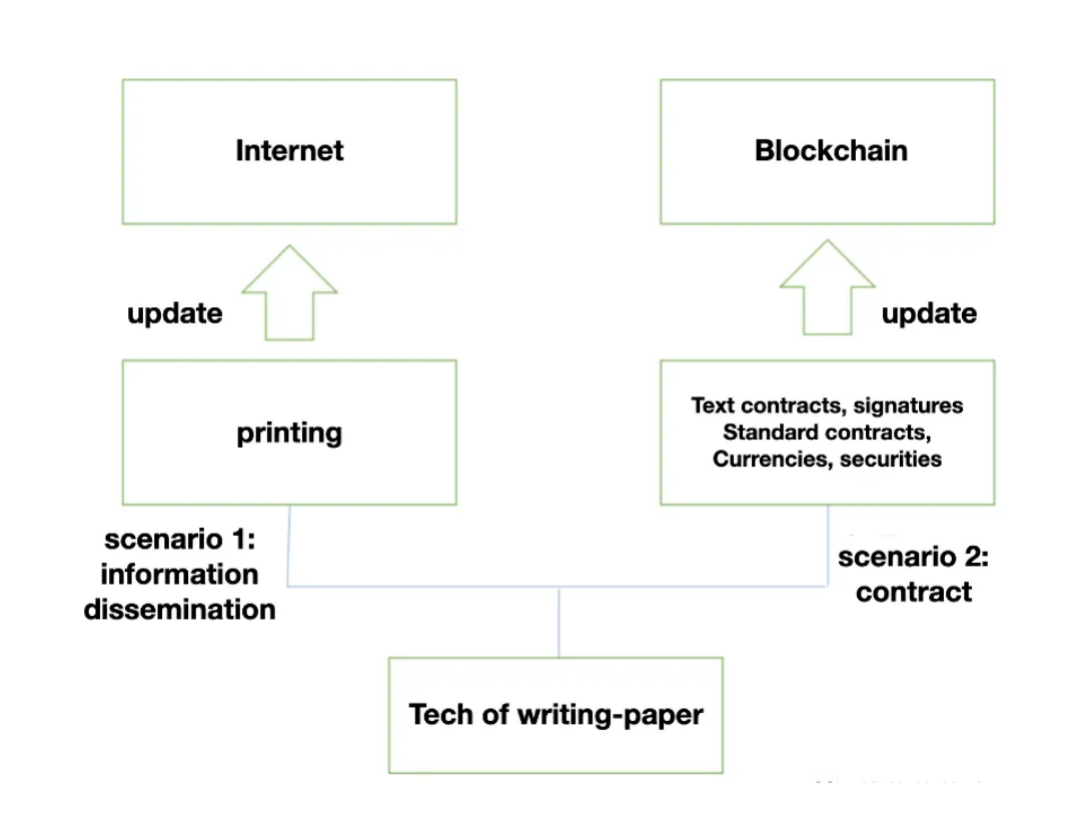}
\caption{Difference between traditional document and blockchain ledger (Image Source: Zhexin Wan, "RWA Tokenization Future Blueprint: Comprehensive Analysis of Underlying Logic and Pathways to Large-Scale Application Implementation", Medium, Feb 26, 2024)}
\label{background}
\end{figure}

The introduction of RWA tokenization has had a significant impact on a variety of industries and is now widely used in both personal asset trading and public asset management. As shown in  Fig.~\ref{background}, similar to the fact that the utilization of the Internet enables the evolution of traditional documents, the application of blockchain is the digital upgrade of text-to-paper technology. 

\subsubsection{Personal Asset Management}
\begin{figure}[htbp]
\center
\includegraphics[width=\columnwidth]{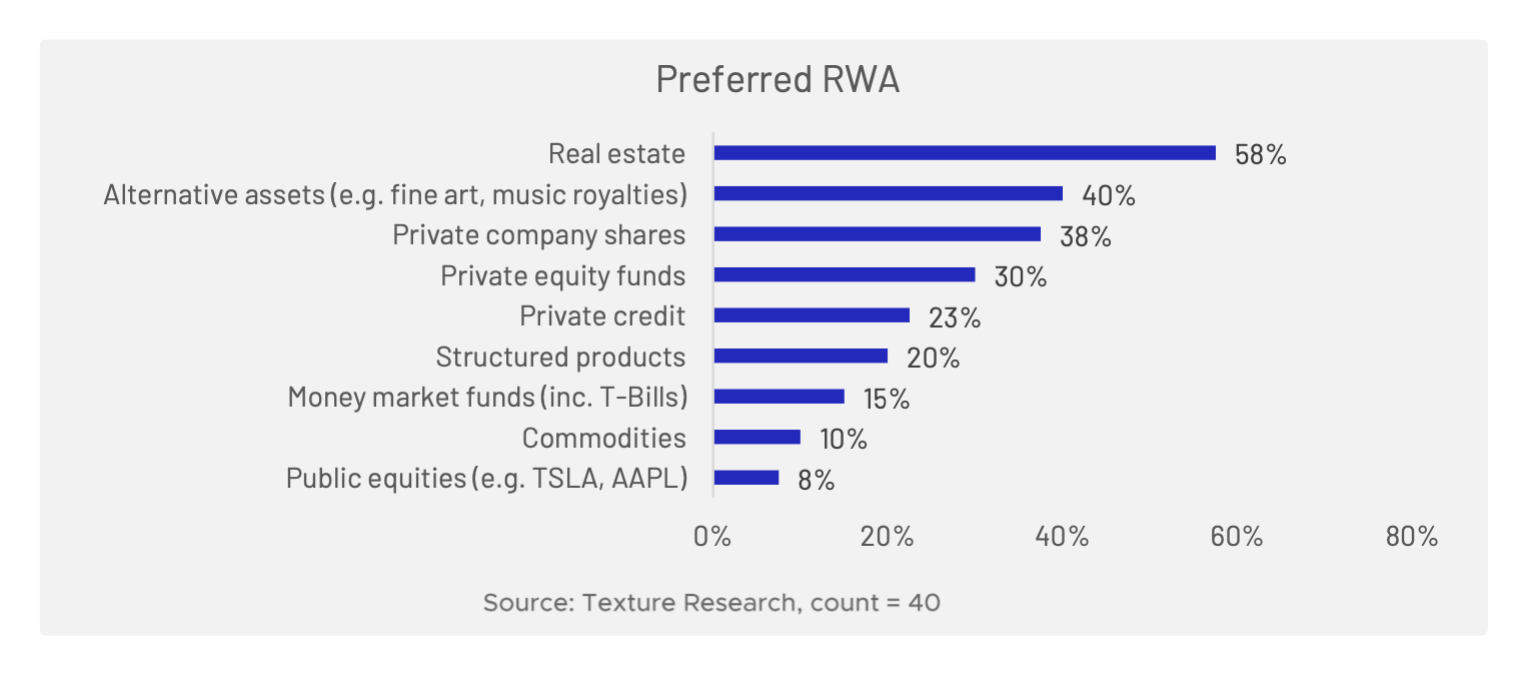}
\caption{Different types of RWA}
\label{RWA}
\end{figure}

Based on the limitations of traditional asset management, the application of tokenizing RWAs allows for increased liquidity and fractional ownership in many fields as shown in Figure ~\ref{RWA}. For example, back in 2018, Elevated Returns (ER) in New York led the world’s first real estate tokenization project. The company utilized Ethereum blockchain technology to sell 18 million digital security tokens, representing ownership shares in the Aspen Ridge Resort \cite{b2}. Additionally, in the art market, traditional trading procedures involve challenges like insufficient transparency and difficult validation of authenticity. Now there are many platforms like Artchain \cite{b3}.  Each masterpiece is marked with an ID, which can be denoted as a token in the smart contract. The ownership of the works can be purchased online while the physical originals of the artworks are stored in a secured warehouse. This guarantees the registration, tracking, protection, and provenance for artworks. 

Applying for patents and trademarks is not only a time-consuming and lengthy process but also comes with high costs. Thus, Non-Fungible Tokens (NFTs) hold significant potential in the field of intellectual property rights. They can enhance transparency and liquidity. While applicants await formal protection from the government, NFTs can offer intellectual property protection. They can also simplify the process of IP application \cite{b4}.

In the commodities market, tokenization can also be adapted in the trading of tangible assets like gold, oil, jewelry, etc. This blockchain solution makes trading more accessible for individuals and enhances the liquidity of the rare products. With the property of security and transparency, blockchain solutions help to establish a novel and fair market environment.

\subsubsection{Public Asset Management}
Countries like Georgia and Sweden have implemented blockchain to manage land registry and asset ownership. In 2016, The official Swedish Land Registry, the Lantmäteriet, developed a platform that leveraged blockchain technology by recording real estate transactions from the moment an agreement to sell is reached until the land title is transferred \cite{b5},\cite{b6}. With this blockchain-based platform, the information asymmetries can be addressed by providing transparency to all involved parties.

In the US, the Bureau of the Fiscal Service's Office of Financial Innovation and Transformation (FIT) has been seeking ways to tokenize grant payments in 2019. In the traditional grant payments allocation process, the grants from the federal government are transferred to a prime recipient, and then to many subrecipients. But this process requires the recipients to submit numerous reports and discussions. Also, the federal government can easily lose visibility and transparency about the payments \cite{b7}. Thus, FIT decided to tokenize the data elements including descriptive information, payment information, and direct or indirect costs in the grant award letter with blockchain technology as evidence. The technique can make it possible to store important data permanently and transfer it between different participants efficiently.

\section{Technical Foundation}
The successful implementation of RWAs tokenization relies heavily on robust technical foundations, primarily anchored in blockchain technology. This section provides an overview of the key technical components essential for understanding the intricacies of RWA tokenization.

\subsection{Blockchain Overview}
Blockchain technology provides a decentralized digital ledger which records transactions across many computers. Thus, the registered transactions can’t be altered retroactively. This technology lays a foundation for cryptocurrency systems and is valuable for the integrity and verifiability of data without a central authority.

\subsection{Consensus Mechanisms}
Consensus mechanisms are protocols that make sure all participants in a blockchain network agree on the current state of the ledger. They are critical to maintaining the security and integrity of the blockchain. Common consensus mechanisms include Proof of Work (PoW) which involves complex computations to secure transactions. While Proof of Stake (PoS) selects validators based on their token holdings. Additionally, Delegated Proof of Stake (DPoS) makes token holders vote for delegates who manage the blockchain. These protocols are fundamental in maintaining the blockchain's security and integrity and crucial for RWA tokenization.

\subsection{Smart Contracts}
Smart contracts are self-executing contracts with the terms of the agreement directly written into lines of code. They automatically enforce and execute the terms of a contract based on its code, which is publicly verifiable on the blockchain. This is instrumental in automating processes, reducing the need for intermediaries, and ensuring transparency and trust.

\section{Standards and Regulatory Considerations}
As the realm of asset tokenization continues to evolve, this section provides insights into the prevailing standards for tokenization and the multifaceted regulatory challenges inherent in this emerging domain.

\subsection{Current Standards for Tokenization}
\subsubsection{ERC-20}
The ERC-20 standard defines a common set of rules that an Ethereum token has to implement\cite{b8}. With this standard, the developers can create new tokens within the Ethereum ecosystem. This token standard became popular due to its simplicity and fundamental role in facilitating the creation of fungible tokens.
\begin{itemize}
    \item \textit{Simplicity}: ERC-20 provides a minimal set of functionalities for tracking and transferring tokens, including key methods like \texttt{transfer}, \texttt{approve}, \texttt{balanceof}, \texttt{allowance} and \texttt{transferFrom}. Thus, ERC-20 tokens are supported by nearly all Ethereum wallets and exchanges, which facilitates user adoption greatly.
    \item \textit{Allowance Mechanism}: This allows token holders to authorize other addresses to spend a certain amount of tokens on their behalf. This provides a secure way to interact with smart contracts without transferring tokens directly.
\end{itemize}
\subsubsection{ERC-721}
ERC-721 defines a standard for non-fungible tokens (NFTs), which is used to implement a standard API within smart contracts for tracking and transferring NFTs\cite{b9}.
\begin{itemize}
    \item \textit{Uniqueness}: Each token is distinct and can have different values even within the same platform. This makes ERC-721 applied for representing ownership of assets such as digital art, collectibles, and real estate in the virtual world.
    \item \textit{Metadata Extension}: Tokens can have associated metadata that can be retrieved via a URI (Uniform Resource Identifier), which can link to details like attributes or images to represent the token. This code provides essential functions for managing metadata URIs in an ERC-721 token contract.
    \item \textit{Enumeration Extension}: This extension provides a way for contracts to enumerate all tokens owned by an address as well as all tokens tracked by a contract. This is an optional function for all ERC-721 contract.
\end{itemize}
\begin{lstlisting}[linewidth=\columnwidth,breaklines=true]
    // A mapping from token ID to URI with metadata
    mapping (uint256 => string) private _tokenURIs;

    // Returns a URL that points to the metadata for the specified token ID
    function tokenURI(uint256 tokenId) public view returns (string memory) {
        require(_exists(tokenId), "ERC721Metadata: URI query for nonexistent token");
        return _tokenURIs[tokenId];
    }

    // Setting the token URI (an internal function)
    function _setTokenURI(uint256 tokenId, string memory uri) internal {
        _tokenURIs[tokenId] = uri;
    }
\end{lstlisting}
\subsubsection{ERC-1155}
ERC-1155 standard expands upon the ideas of ERC-20 and ERC-721. It’s particularly useful in gaming and decentralized finance where a user may perform transactions involving multiple token types\cite{b10}.
\begin{itemize}
    \item \textit{Flexibility}: Supports fungible, non-fungible, and semi-fungible tokens in a single contract.
    \item \textit{Batch Transfers}: Supports multiple types of tokens to be transferred simultaneously, which can reduce transaction costs significantly when dealing with multiple token transfers.
\texttt{safeBatchTransferFrom} function in the ERC-1155 standard allows for the simultaneous transfer of multiple token types in one transaction. This function checks if the caller is authorized, updates the token balances for both sender and receiver, emits a transaction event, and ensures the recipient can handle the tokens. This streamlined process reduces transaction costs and enhances efficiency for managing diverse assets.
\end{itemize}
\begin{lstlisting}[linewidth=\columnwidth,breaklines=true]
// Enable batch transfers for multiple token types.
function safeBatchTransferFrom(
    address _from,
    address _to,
    uint256[] memory _ids,
    uint256[] memory _values,
    bytes memory _data
) public {
    require(_from == msg.sender || isApprovedForAll(_from, msg.sender),
        "ERC1155: caller is not owner nor approved");
    for (uint256 i = 0; i < _ids.length; ++i) {
        uint256 id = _ids[i];
        uint256 value = _values[i];

        balances[id][_from] = balances[id][_from].sub(value);
        balances[id][_to] = balances[id][_to].add(value);
    }
    emit TransferBatch(msg.sender, _from, _to, _ids, _values);
    _doSafeBatchTransferAcceptanceCheck(msg.sender, _from, _to, _ids, _values, _data);
}
\end{lstlisting}

\subsection{Regulatory Challenges and Solutions}
\begin{enumerate}
    \item \textit{Varied Jurisdictional Regulations}: Different regions exhibit diverse approaches to the regulation of blockchain technologies and tokenized assets. For instance, the U.S. Securities and Exchange Commission (SEC) has been actively engaging in discussions around whether certain tokens qualify as securities, which would subject them to additional regulatory scrutiny and compliance requirements. Similarly, the European Union is working on frameworks like the Markets in Crypto-Assets (MiCA) to streamline regulatory approaches across its member states.
    \item \textit{Compliance and Legal Challenges}: Token issuers must navigate a complex landscape of compliance issues, including but not limited to AML (Anti-Money Laundering) and KYC (Know Your Customer) protocols, and securities regulations. Ensuring compliance requires robust legal expertise and often, substantial resources, especially for platforms operating across multiple jurisdictions.
    \item \textit{Impact on Tokenization and Investor Protection}: Regulations are intended to protect investors by ensuring that token issuers disclose adequate information about the risks and details of the investment, maintain fair trading practices, and operate within the bounds of the law. These protections help stabilize the market and build trust among investors, which is crucial for the adoption and growth of tokenization technologies.
\end{enumerate}

\section{Tokenization Process}

\subsection{Step-by-Step Conversion}
The tokenization of RWAs into digital tokens is a multifaceted process that transforms physical and intangible assets into tradable digital representations on a blockchain. This process begins with the identification and selection of the asset to be tokenized. The chosen asset could be anything from a piece of real estate, or a work of art, to shares in a private company. 

Following the selection of the asset, the asset undergoes a thorough valuation to determine its suitability for tokenization. This valuation not only determines the market value of the asset but also aids in defining the structure of the tokenization—how many tokens represent the asset and at what price each token should be offered. For tangible assets like real estate, this might involve a detailed property inspection and market comparison, whereas for assets like art, it might require authentication and historical significance evaluation.

The next step is the design and development of the token itself, which involves choosing the right blockchain platform that supports token standards such as ERC-20 or ERC-721 and deciding on the type of token—whether it should be fungible (identical and interchangeable) or non-fungible (unique). The token design includes defining what rights and benefits the token holders will have. For example, in the case of a building, this might include not only ownership shares but also potentially a share in the rental income.
After the token design, the creation of a smart contract is imperative. This self-executing contract with the terms of the agreement written into code governs the functionality of the token. It defines the rules under which the tokens operate, including how they can be bought, sold, or transferred. The smart contract automates the execution of these rules, ensuring efficiency and compliance without manual intervention.

With the smart contract in place, the actual issuance of the tokens can occur. This involves the minting of tokens on the blockchain, which are then ready to be distributed to investors or purchasers. This step marks the transformation of the physical or intangible asset into a digital form that can be traded on blockchain platforms.

The final step in the tokenization process is the distribution of the tokens. Depending on the nature of the asset and the target investors, these tokens can be sold directly to known investors, offered through a public sale, or listed on digital asset exchanges for trading. This phase is critical as it not only involves the initial sale of the tokens but also sets up the framework for secondary trading, providing liquidity and accessibility to investors.

Here is an example token following the ERC-20 standard through an Open Zeppelin template.

\begin{lstlisting}[linewidth=\columnwidth,breaklines=true]
pragma solidity ^0.8.24;

import "./ERC20.sol";

contract MyToken is ERC20 {
    constructor(string memory name, string memory symbol, uint8 decimals)
        ERC20(name, symbol, decimals)
    {
        // Mint 100 tokens to msg.sender
        // Similar to how
        // 1 dollar = 100 cents
        // 1 token = 1 * (10 ** decimals)
        _mint(msg.sender, 100 * 10 ** uint256(decimals));
    }
\end{lstlisting}

\subsection{Smart Contracts Role}
Smart contracts are foundational to the functionality and security of blockchain technology, particularly in the context of asset tokenization. These autonomous contracts, once deployed on the blockchain, execute automatically when predefined conditions are met, ensuring the seamless management of digital tokens without the need for a centralized authority. Their capabilities revolutionize the way transactions are conducted and ownership rights are enforced in the realm of decentralized finance \cite{b11}.

In the realm of ownership management, smart contracts are pivotal. They maintain a precise and immutable record of ownership on the blockchain. Every token has its ownership details coded into the smart contract, making each token's provenance and ownership transparent and indisputable \cite{b12}. This level of transparency is crucial in markets where trust is paramount, eliminating the need for lengthy verification processes traditionally handled by third parties such as banks or legal teams.

When it comes to the transfer of tokens, smart contracts automate what was once a cumbersome and error-prone process \cite{b11}. Upon the execution of a transaction, the smart contract immediately updates the blockchain ledger with new ownership details. This automation ensures that transfers are not only secure but occur instantaneously, thereby enhancing liquidity and enabling real-time trading \cite{b13}. This feature is particularly transformative in markets like real estate or art, where sales and transfers traditionally involve multiple steps and can take considerable time to finalize.

Furthermore, in scenarios where tokens represent shares in a revenue-generating asset, such as rental properties, smart contracts facilitate the distribution of dividends \cite{b12}. They can be set to automatically distribute earnings to token holders according to their ownership stake, ensuring fairness and timeliness in profit-sharing. This feature not only simplifies the management of such assets but also enhances their attractiveness to investors by providing a clear, automated, and trustless dividend distribution mechanism.

Overall, the role of smart contracts in managing the ownership and transfer of tokens is transformative, offering unprecedented levels of efficiency, security, and compliance in the digital asset sphere. These contracts are the backbone of the decentralized finance (DeFi) ecosystem, providing the tools necessary to reshape the landscape of investment and asset management.

\begin{lstlisting}[linewidth=\columnwidth,breaklines=true]
pragma solidity ^0.8.4;

contract AssetTokenization {
    address public owner;
    mapping(address => uint) public ownershipShares;

    event OwnershipTransferred(address indexed previousOwner, address indexed newOwner);
    event SharesIssued(address indexed to, uint shares);

    constructor() {
        owner = msg.sender;
    }

    modifier onlyOwner() {
        require(msg.sender == owner, "Only the owner can perform this action");
        _;
    }

    function transferOwnership(address newOwner) public onlyOwner {
        require(newOwner != address(0), "Invalid owner address");
        emit OwnershipTransferred(owner, newOwner);
        owner = newOwner;
    }

    function issueShares(address to, uint shares) public onlyOwner {
        require(to != address(0), "Invalid address");
        ownershipShares[to] += shares;
        emit SharesIssued(to, shares);
    }

    function transferShares(address from, address to, uint shares) public {
        require(ownershipShares[from] >= shares, "Insufficient shares");
        require(to != address(0), "Invalid address");

        ownershipShares[from] -= shares;
        ownershipShares[to] += shares;
    }
}
\end{lstlisting}
In this example, the smart contract defines a basic framework for asset tokenization:
\begin{itemize}
    \item The owner of the contract can issue shares (issueShares) to any address. The owner can transfer the contract's ownership to another address using the transfer ownership function.
    \item Shareholders can transfer their shares to others with the transferShares function.
\end{itemize}

This simplistic example illustrates how a smart contract can be used to manage ownership and transfer rights within the framework of tokenized assets. 

\section{Advantages of RWA Tokenization}
\begin{figure}[htbp]
\center
\includegraphics[width=\columnwidth]{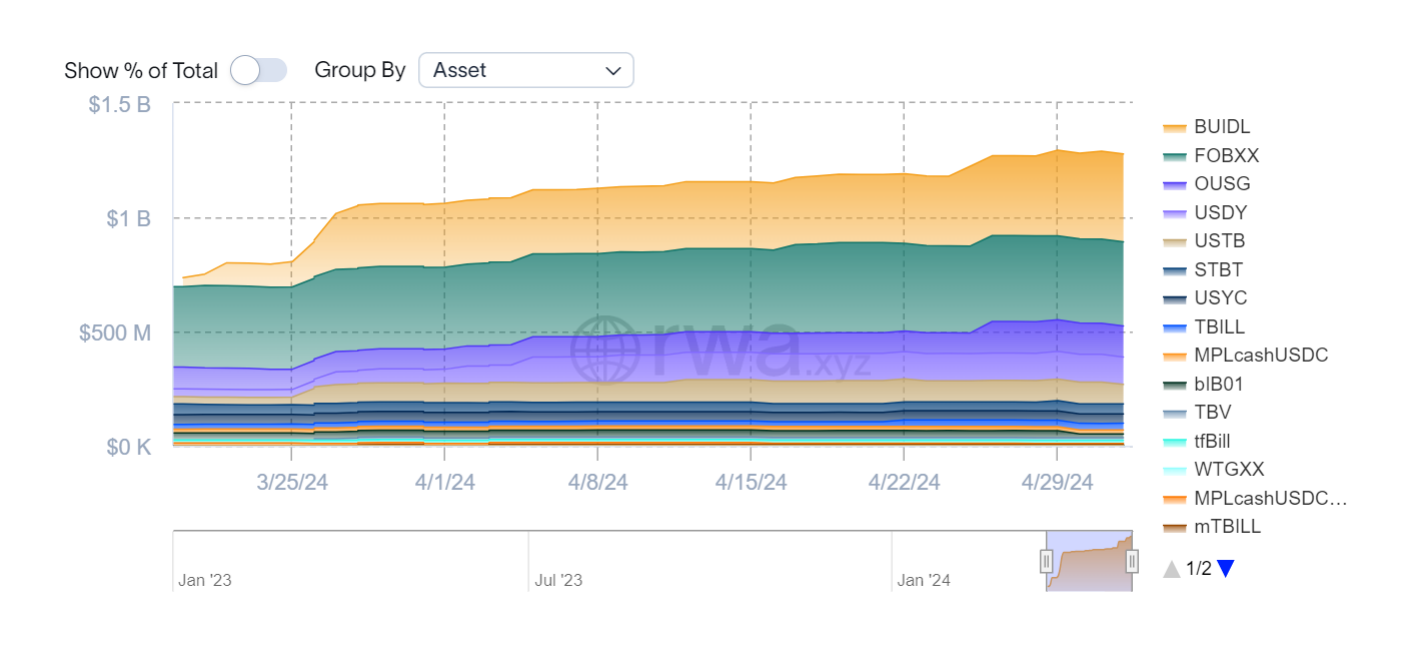}
\caption{The Value of Tokenized Treasury Market in the U.S. (Data Source: Tokenized Treasuries market (RWA.xyz))}
\label{treasure}
\end{figure}

According to the data from the tokenized RWA statistics platform RWA.xyz in Figure ~\ref{treasure}, the value in the tokenized treasury sector has soared to nearly \$1.3 billion in April 2024 from \$100 million at the start of 2023. This significant increase indicates an urgent need for a more advanced method to manage RWAs, including asset ownership, trading monitoring, and asset origin tracing.

\subsection{Enhanced Liquidity}
The primary advantage of RWA tokenization is that through digitizing asset ownership, the RWA assets can be divided into smaller, more easily tradable units. Thus, compared the traditional entire ownership trading, the ownership can be divided. This property of fractionalization enhances liquidity because it lowers the entry barrier of the market and many partially qualified investors and buyers. Additionally, asset trading, involving the exchanging of digitalized tokens, is much easier because there is no need to submit redundant paper documents. The circulation on the secondary market makes the prices transparent to other sellers and buyers in this field, thus providing the platform to compare prices and stimulate market activity.

Unlike the traditional asset transaction, with the technique of RWA tokenization, various kinds of assets can be represented with digital tokens regardless of the location, form, or physical properties of the assets. The transaction with digital tokens overcomes the limitation of trading hours of the stock exchanges. As the transaction does not require the validation of an authenticated third party, the buyers and sellers can initiate a trade whenever they want. This is especially beneficial to the across-broader transaction in different time zones. This flexibility brings about the advantages of mitigating risks like geographic distance, shipment, and time differences.

\subsection{Increased transparency}
In traditional asset management, especially in real estate transactions, it is often a tedious process to investigate any existing contracts, leases, or accuse associated with the property \cite{b14}. With a blockchain-based registry, these details could be recorded and updated transparently. All relevant information about the property’s legal status, previous ownership, and fundamental facts can be readily available on the blockchain. So the buyers can obtain sufficient information without extensive legal searches and document reviews.

Similarly, in commodity markets, the application of RWA tokenization provides a detailed document of the managing, tracking, and verification of assets. For instance, the blockchain can be utilized to authenticate the origin of rare jewelry, tracing from mining, cutting, and wholesale to retailing, which reduces the information asymmetry.

\subsection{Improved security in asset management}
RWA tokenization can also improve security during asset registration and exchange. As traditional asset management often relies on physical registration certificates, the certificates can face risks of loss, theft, or destruction. Even if there may be digitized copies of the certificates, they may not be updated among different bureaucracies in different countries.  The host servers are also vulnerable to risks of being hacked or destroyed. For instance, in 2010, an earthquake in Haiti destroyed the host server of the land registry database. This disaster led to a loss of sixty years' worth of land registry records. Over a million citizens who suffered property loss in this disaster were unable to prove their assets \cite{b15}. In these cases, the property of immutability of RWA tokenization based on blockchain technology can help. It ensures that even if some nodes in the blockchain are destroyed, the complete ledger can still be kept because at least other intact full nodes also store copies of the data \cite{b16}. Therefore, the integrity and consistency of the data can be guaranteed.

During the trading process of the RWA, the security of the transaction can also be ensured even without an authenticated third party. The ownership of a property is represented by digital tokens recorded on the blockchain ledger. Only the legitimate owner of the asset can initiate a transaction due to the private key validation. Once the transaction is validated, it cannot be immutable. This process automatically avoids the risk of double-spending and fraud, adding an additional layer of trust to the transaction. 

\subsection{Reduced transaction cost}
In traditional asset transactions, a legal trade process should involve the witness of intermediaries, like brokers, agencies, custodians, and so on. Their validation and paperwork can lead to additional high transaction costs. As tokenized RWA simplifies the whole transaction process and makes point-to-point trading possible on blockchain platforms, the transaction costs can be largely reduced. According to the research and estimation of Roland Berger in Figure 4, with the application of tokenized RWA, about €4.6 billion of the transaction costs in equity trading could be saved by 2030. This estimation only covers one asset type\cite{b17}.  So in the general scenario, the cost savings will be significantly much greater than the expectation. 

\begin{figure}[htbp]
\center
\includegraphics[width=\columnwidth]{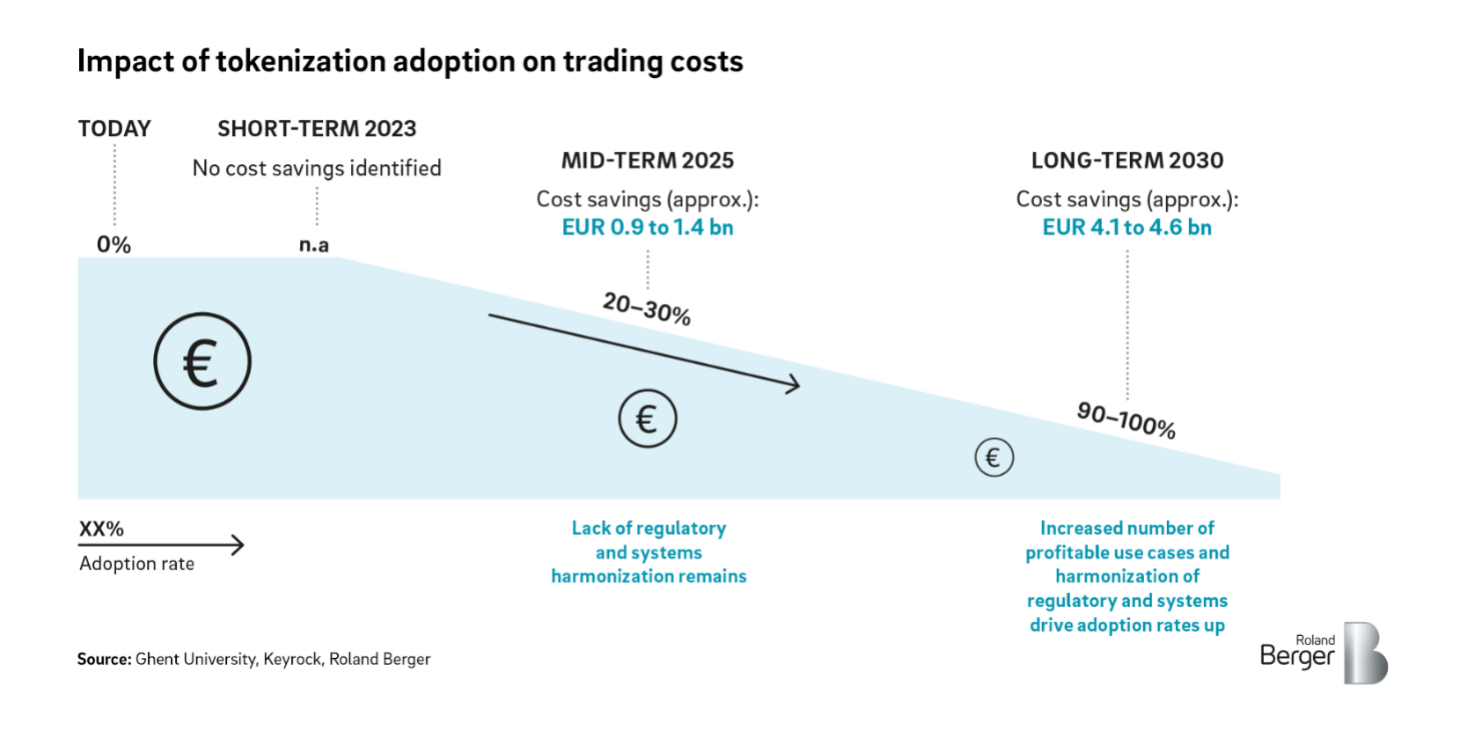}
\caption{Impact of tokenization adoption on equity trading costs}
\label{impact}
\end{figure}

\section{Case Studies}

\subsection{Real estate}
Real estate is defined as the land and any permanent structures, like a home, or improvements attached to the land, whether natural or man-made. Real estate is a form of real property. It differs from personal property, which is not permanently attached to the land, such as vehicles, boats, jewelry, furniture, and farm equipment.

\subsubsection{Technical Details}
\begin{itemize}
    \item \textit{Blockchain Infrastructure}: Real estate tokenization leverages blockchain technology, ensuring transparent, secure, and immutable records of property ownership and transactions. Smart contracts automate processes like ownership transfer and rental agreements, enhancing efficiency. 
    \item \textit{Token Standards}: Real estate assets are tokenized using specific standards like ERC-20, ERC-721, or proprietary protocols. These standards define attributes such as fractional ownership, voting rights, and revenue-sharing mechanisms.
    \item \textit{Compliance and Legal Frameworks}: Tokenization platforms adhere to regulatory requirements, ensuring compliance with securities laws, property regulations, and Know Your Customer (KYC) procedures. Compliance is crucial for investor protection and market integrity.
\end{itemize}

\subsubsection{Financial Details}
\begin{itemize}
    \item \textit{Liquidity and Accessibility}: Real estate tokenization enhances liquidity by fractionalizing property ownership, allowing investors to buy and sell tokens in secondary markets. This increases accessibility to real estate investment opportunities for retail investors and reduces barriers to entry.
    \item \textit{Diversification and Risk Management}: Investors can diversify their portfolios by investing in fractionalized real estate assets across different locations, types, and risk profiles. Tokenization enables risk management through asset diversification and liquidity enhancement.
    \item \textit{Income Generation}: Real estate tokens may generate income for investors through rental yields, property appreciation, or revenue-sharing agreements embedded in smart contracts. This creates passive income streams and enhances the attractiveness of real estate investment.
\end{itemize}

\subsubsection{Operational Details}
\begin{itemize}
    \item \textit{Asset Management}: Tokenization platforms facilitate property management tasks such as rent collection, maintenance, and tenant screening. Smart contracts automate these processes, reducing administrative overhead and ensuring transparency.
    \item \textit{Marketplaces and Exchanges}: Digital marketplaces and exchanges facilitate the trading of real estate tokens, providing liquidity and price discovery mechanisms. These platforms enable investors to buy, sell, and trade tokens seamlessly.
    \item \textit{Governance and Decision-Making}: Token holders may participate in governance processes through voting mechanisms embedded in smart contracts. This allows stakeholders to make decisions on property management, asset upgrades, and strategic initiatives.
\end{itemize}

\subsection{Trading Cards}
NFT Trading cards are digital representations of collectible cards that are owned and traded on blockchain platforms. They leverage the unique properties of NFT’s such as scarcity, provable ownership, and immutability, to offer collectors a new way to engage with their digital assets.

\subsubsection{Technical Details}
\begin{itemize}
    \item \textit{Blockchain Technology}: NFT trading cards are built on blockchain technology, which provides a secure, transparent, and immutable ledger for recording ownership and transaction history.
    \item \textit{Token Standards}: NFT trading cards adhere to specific token standards like ERC-721 or ERC-1155, which define the characteristics and functionalities of non-fungible tokens, including uniqueness, ownership, and transferability.
    \item \textit{Metadata and Interoperability}: Each NFT trading card contains metadata that describes its attributes, such as artwork, rarity, edition, and creator information. Interoperability standards allow NFTs to interact with other platforms and applications, enhancing their utility and value.
\end{itemize}

\subsubsection{Financial Details}
\begin{itemize}
    \item \textit{Market Dynamics}: The value of NFT trading cards is driven by factors such as scarcity, demand, reputation of creators, and historical sales data. Rare or exclusive cards often command higher prices in the secondary market.NFT trading cards enable various monetization models, including primary sales, secondary market transactions, royalties for creators, licensing agreements, and brand partnerships.
    \item \textit{Investment Opportunities}: Investors can diversify their portfolios by investing in NFT trading cards, leveraging the potential for appreciation in value over time. However, the market is subject to volatility and speculation risks.
\end{itemize}

\subsubsection{Operational Details}
\begin{itemize}
    \item \textit{Digital Ownership Experience}: NFT trading cards offer collectors a novel digital ownership experience, allowing them to display, trade, and interact with their cards in virtual environments, such as digital galleries or gaming platforms.
    \item \textit{Marketplaces and Platforms}: Digital marketplaces and platforms facilitate the buying, selling, and trading of NFT trading cards, providing liquidity and exposure to a global audience of collectors and enthusiasts.
    \item \textit{Community Engagement}: Vibrant communities of collectors, artists, developers, and enthusiasts contribute to the growth and sustainability of the NFT trading card ecosystem through collaboration, networking, and cultural exchange.
\end{itemize}

\section{Challenges and Limitations}
The promising landscape of blockchain technology and the tokenization of RWAs are not without their challenges and limitations, which can impede progress and implementation on a larger scale.

\subsection{Scalability}
One of the most significant technical challenges facing blockchains, especially those like Ethereum which are popular for tokenization, is scalability. The ability to process transactions quickly and cost-effectively is crucial as the volume of transactions on blockchain networks increases. Currently, networks such as Ethereum can become congested, leading to slower transaction times and higher costs. This limitation can be a significant barrier to the widespread adoption of blockchain for asset tokenization, where speed and efficiency are critical \cite{b18}.

\subsection{Immutability}
While blockchain technology enhances security, it also introduces new vulnerabilities, particularly in the rigid structure and execution of smart contracts. Unlike traditional contracts, which allow for amendments and adaptations to address real-life dynamics, smart contracts are inflexible once deployed, potentially leading to practical complications if changes are necessary. This rigidity can expose tokenization platforms to increased risks as programming errors or oversights become more challenging to rectify post-deployment \cite{b13}.  The result is that amending a smart contract may not only complicate the modification process but also increase transaction costs and the likelihood of errors in reflecting intended changes accurately \cite{b19}. As tokenization platforms grow more complex, the potential for such vulnerabilities increases, necessitating rigorous security audits and updates to ensure the integrity of the platforms.

\section{Future Perspectives and Innovations}

\begin{figure}[htbp]
\center
\includegraphics[width=\columnwidth]{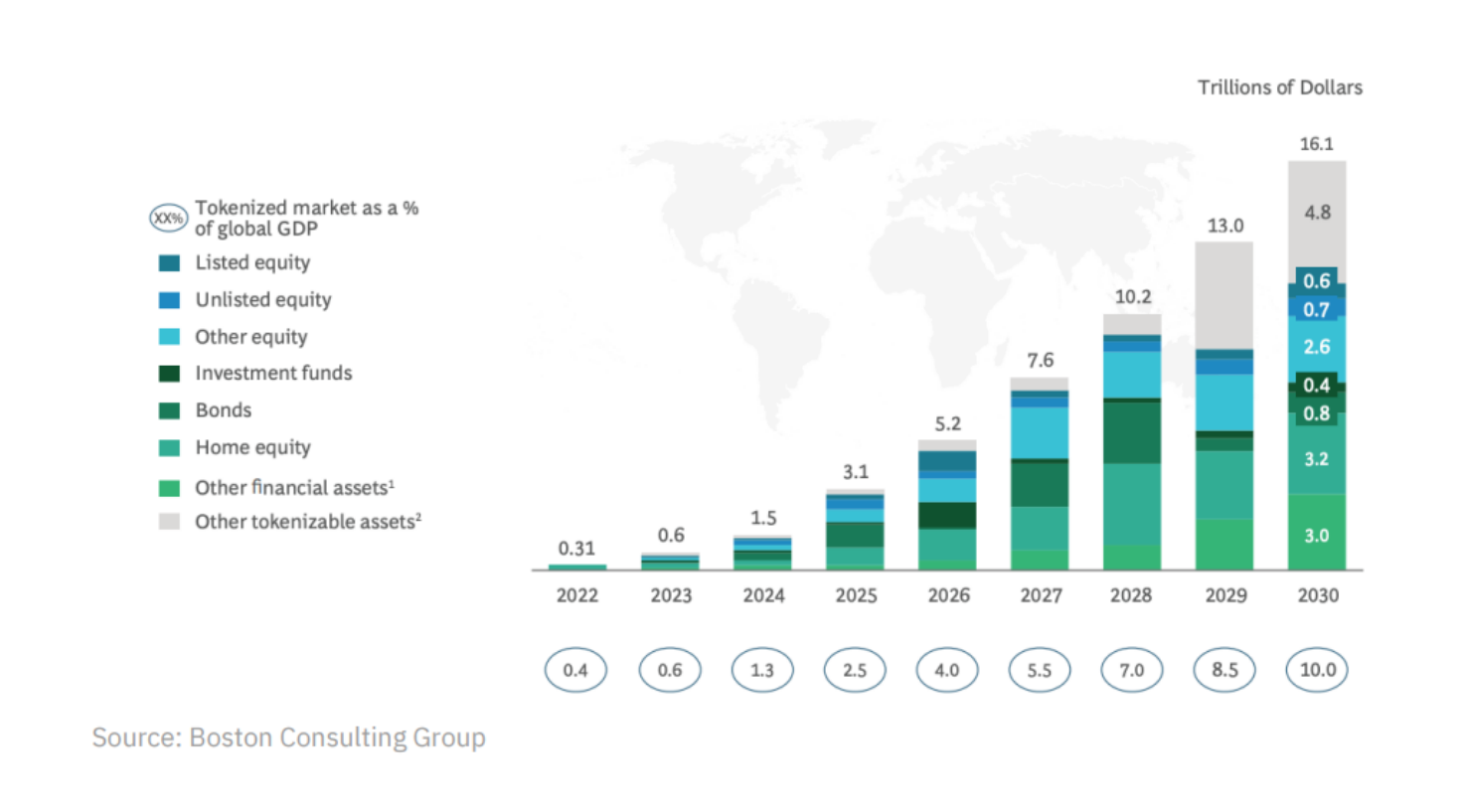}
\caption{Impact of tokenization adoption on equity trading costs}
\label{2030}
\end{figure}

Despite the challenges, the future of blockchain and its application in the tokenization of RWAs is bright. According to Binance’s July 2023 RWA market research, The market for tokenized assets is projected to grow substantially, reaching an estimated \$16 trillion by 2030 \cite{b20}. As tokenization of RWAs continues to develop, several promising trends are emerging in the industry:

\subsection{Interoperability and Cross-Chain Transactions}
Cross-chain capabilities allow for tokenized assets to be traded, utilized, or moved freely among different blockchain environments, which further enhances liquidity and accessibility \cite{b18}. This expansion may facilitate a worldwide ecosystem for tokenized RWAs and minimize fragmentation within the ecosystem \cite{b21}.

\subsection{New Markets}
The potential for tokenizing new types of assets is vast, including areas such as intellectual property, personal data, or carbon credits, which could transform a multitude of industries \cite{b21}. For example, tokenizing intellectual property could streamline the licensing processes, making it easier for creators to monetize and protect their work. Similarly, personal data tokenization could give individuals more control over their data and the ability to monetize their personal information securely and transparently.

\subsection{Impact on Financial Systems}
Widespread tokenization has the potential to dramatically reshape global financial markets and investment strategies. By making assets more liquid and accessible, blockchain could democratize access to investment opportunities, previously available only to select groups or individuals. Moreover, the increased transparency and efficiency provided by blockchain could lead to more robust and resilient financial systems, reducing the likelihood of fraud and increasing the ease of regulatory compliance.

\section{Conclusion}
In conclusion, the advent of blockchain technology has ushered in a new era of asset management through tokenization, particularly evident in RWAs like real estate and NFT trading cards. This paper has elucidated the technical, financial, and operational details of RWA tokenization, highlighting its potential to revolutionize traditional asset ownership and investment paradigms.

By leveraging blockchain infrastructure, token standards, and smart contracts, RWA tokenization offers enhanced liquidity, accessibility, and transparency. It addresses longstanding challenges in asset management, such as high entry barriers, limited liquidity, and cumbersome transaction processes. Moreover, RWA tokenization facilitates income generation, risk management, and community engagement, fostering a vibrant ecosystem of investors, creators, and enthusiasts.

Looking ahead, the widespread adoption of blockchain tokenization in asset markets holds promise for significant societal and economic impacts. If blockchain tokenization becomes mainstream, it could democratize access to investment opportunities, catalyze economic growth, and promote financial inclusion. However, realizing these benefits requires collaboration among stakeholders, regulatory clarity, and ongoing innovation.

\section*{Contribution}
All authors discussed the results and contributed to the final manuscript. 

Yixuan Li wrote background knowledge of the abstract, introduction, project framework, intermediate parts, and conclusion, combined all parts together, and included references.

Yimin Yang collected background information and previous researches, summarizing to the introduction and motivation, and the advantages of RWA tokenization. 

Xiaolei Zhao was responsible for the technical foundation as well as standards and regulatory considerations critical to compliance and performance.

Ning Xia conducted the tokenization process step by step, analyzed the critical role of smart contracts, and discussed the limitations and future perspectives of RWAs and Tokenization. 

Yucong Li completed the case study and analyzed technical, financial, and operational details.

\end{document}